# Unsourced Sparse Multiple Access: a promising transmission scheme for 6G massive communication

Yifei Yuan*, Yuhong Huang*, Chunlin Yan[+], Sen Wang*, Shuai Ma*, and Xiaodong Shen*

*China Mobile Research Institute, [+] ZGC Institute of Ubiquitous-X Innovation and Applications

## ABSTRACT

**Massive communication is a key scenario of 6G which requires a hundred times higher connection density compared to 5G. As a promising direction, unsourced multiple access has a capacity bound much higher than those of orthogonal multiple access or slotted-ALOHA. In this paper we describe a framework of unsourced sparse multiple access (USMA) realization that consists of two key modules: compressed sensing for preamble generation and sparse interleaver division multiple access for main packet transmission. By proper combination of various components such as sensing matrix, channel codes, interleaver, receiver algorithms, USMA offers a more feasible engineering design that can support larger number of users than the state-of-the-art scheme. To illustrate the scalability of USMA, a customized implementation is proposed for ambient Internet of Things to further reduce the memory and computation complexities. Simulations results of Rayleigh fading and realistic channel estimation show that the solution can deliver nearly 4 times resource efficiency as of traditional slotted-ALOHA systems.**

## I. INTRODUCTION

As one of six usage scenarios, massive communication is included in the framework and overall objectives of 6G by International Telecommunication Union [1]. Compared to the connection density requirement in 5G massive machine type communication (mMTC) which is $10^6$ devices/km$^2$, massive communication of 6G has a much higher target, ranging from $10^6$ to $10^8$ devices/km$^2$. There is a wide range of use cases and applications for massive communication, one of which is ambient Internet-of-Things (A-IoT) with ultra-low complexity and power consumption. The strong interests propel a study item approved by the Third Generation Partnership Project (3GPP) which focuses on the technical solutions and performance evaluations of A-IoT [2].

For 5G mMTC, non-orthogonal multiple access (NOMA) was studied as a candidate technology to improve the spectral efficiency [3]. However, no subsequent standardization work was pursued, e.g., the physical layer of 5G mMTC is still orthogonal multiple access (OMA) based. In light of 10~100 times higher connection density for 6G massive communication, new multiple access is expected to play important role in meeting these requirements. Consequently, in IEEE Communications Society, an Emerging Technology Initiative (ETI) on next generation multiple access was established in 2021. In IMT-2030 (6G) Promotion Group of China, a task force group for new multiple access was created in 2022.

In massive connection scenario, control signaling overhead and power consumption of devices are the main bottlenecks. They can be substantially reduced by unsourced multiple access (UMA), a promising approach where random access and grant-free small-package transmission are inherently integrated. In UMA, no user-specific signatures are allocated prior to the transmissions. A common codebook is shared by all potential users and each active user can randomly select a codeword to encode its data. Recently, there were some breakthroughs in information theoretic analysis where capacity bounds were derived for UMA [4]. The significant performance potential of UMA inspired wide interests in its designs [5-8] which also include new channel coding schemes [9][10] and tree search-based compressed sensing (CS) approach [11]. In [12], a framework of UMA was introduced which consists of two basic modules: CS-based sequence generation and sparse interleaver division multiple access (SIDMA).

In this paper we will elaborate the designs of USMA and demonstrate its practicality, cost-effectiveness and scalability by various parameters and algorithm optimizations, supported by extensive simulations. The effort is important and timely since 3GPP will kick off its 6G study in June 2025 where technical solutions to massive communication are expected to be a key research area.

The paper is organized as follows. In Section II, scenarios, performance requirements and technical challenges of massive communication are discussed. The design, called unsourced sparse multiple access (USMA) is described in Section III with simulation results. The scalability of USMA, especially for ambient IoT, is elaborated in Section IV with customized design, parameters and simulation results. Finally, the paper is concluded by Section V.

## II. REQUIREMENTS AND DESIGN CHALLENGES

### A. General requirements

It is envisioned that 6G massive communication would feature a variety of IoT devices. These devices would support expanded and new applications in smart cities, transportation, logistics, health, energy, environmental monitoring, agriculture and etc. While high connection density and low power consumption are the general requirements of 6G massive communication, different applications have their unique requirements.

The level of difficulty to support massive connection depends heavily on traffic models. If the traffic is deterministic with long period, radio resources can be pre-allocated, so that different users can be orthogonally multiplexed. However, in many applications, traffics are sporadic short packets. Given the massive number of such devices (e.g., 10~100/m$^2$), there is a chance that many of them would try to access systems simultaneously.

Mobility support is important since 6G massive communication should serve more diverse applications such as sensors or tags on moving vehicles or livestock roaming on the farms, satellite communication networks. Mobility makes it more difficult to pre-allocate resources or identity related information, meaning that the system needs to efficiently deal with a large number of users trying to randomly access. The channels may also be time-varying, causing fast fading.

Many IoT devices would be installed in basement or behind thick walls. The pathloss can be well beyond 110 dB. Hence, the waveforms, modulation, coding should be designed for very low signal to noise ratio (SNR). To combat the severe pathloss, lower bands, e.g., < 1 GHz are often used. In such a case, due to the longer wavelength compared to the mid-band or millimeter band, the number of antennas that can be installed on a base station is typically 2 or 4, rather than 32 or 64. Hence, less degree of freedom can be exploited in the spatial domain.

Transmission delay includes buffer delay and over-the-air latency. Buffer delay depends primarily on the transmission scheme. If random access and data transmission are combined into a single transmission, e.g., two-step random access channel (2-step RACH), buffer delay can be small. However, in the current standards, there is no effective physical layer solution to handle the collisions in 2-step RACH. Over-the-air latency is largely determined by the packet size and data rate, e.g., operating SNR. It can be long if devices are in poor coverage. For many IoT applications, the required transmission delay should be at the level of hundred millisecond to second.
.

### B. Ambient IoT

Ambient IoT is a type of massive communication with several unique characteristics. Many ambient IoT devices do not have batteries, e.g., they harvest the electromagnetic energy radiated by feeders. A feeder can be a base station, a node or a radio frequency identification (RFID) reader that can power ambient IoT devices by transmitting carrier waves (CW). For ambient IoT devices without battery, there are two sub-classes: Deice 1a without any energy storage and Device 1b with small energy storage. Device 1a can only transmit signals when it is "illuminated" by CW. Because of this, Device 1a is also called backscatter device, e.g., to modulate the CW and "reflect" it back. Device 1a can be massively manufactured with extremely low cost, e.g., about 1 cent per device. However, the lack of energy storage limits the "transmit" power of Device 1a and thus its coverage. The low-cost requirement puts significant constraints on the signal processing capability as well as the memory size. In contrast, Device 1b can harvest the energy from CW even when it is not transmitting. Hence, its transmitting power can be higher, leading to the wider coverage. The cost pressure on Device 1b is less compared to Device 1a. Hence, more advanced signal processing can be supported. Nevertheless, Device 1b is noticeably more expensive than Device 1a, under the current manufacturing process.

**Table 1** Key performance indicators and deployment parameters of inventory scenario of Ambient IoT.

| Parameter | Value |
|---|---|
| Maximum transmission latency | 1 second |
| Range of coverage | 30 m |
| Data rate | < 1 kbps |
| Message size | < 256 bits, e.g., 96 bits |
| Connection density | 1.5 million per km$^2$ |
| Mobility speed | < 6 km/h, typically 3 km/h or less |
| Peak inventory rate | 100 devices per second |
| Operating bands | 900 MHz, FDD |
| System bandwidth | 180 kHz |
| Number of antenna elements at reader | 1 or 2 |
| Block error rate (BLER) target | 5% |

In 3GPP, 32 use cases of ambient IoT are identified. There use cases can be further grouped into four types: warehouse inventory, environment sensing, terminal positioning, and control commands. From the connection density perspective, indoor warehouse inventory is the most challenging, compared to three other types. Inventory automation is also in most urgent demand, and thus becomes the first priority in [2]. Table 1 lists several key performance indicators and deployment parameters of inventory scenario of ambient IoT which help to guide the physical layer design.

### C. Design challenges of uplink massive communication

From the discussions in Section II-A and II-B, it can be seen that the designs of uplink massive communication have the following challenges.

*1) Low cost and low power consumption of devices*

These are the prerequisite of massive deployment, meaning that the signal process capability and memory size are very limited in the device transmitters. The signaling exchange between the network and devices should be kept minimum.

*2) Supporting a number of devices simultaneously*

Because of random nature of traffic arrivals and massive number of devices, the number of simultaneous transmissions in the same time and frequency resources can be high. This is exacerbated by the limited spatial dimensions in the channels.

*3) Small packet size in multiple access channel*

Channel coding schemes of enhanced mobile broadband (eMBB) are usually designed for large block sizes in additive white Gaussian noise (AWGN) channel. However, for many IoT applications, the packet sizes are quite small, e.g., around 100 bits. In addition, when superposition transmission is supported, cross-user interference would be dominant rather than the thermal noise. In this case, channel codes optimization is still an open problem.

## III. DESIGN FRAMEWORK

As discussed in Section II, the physical layer design of massive communication should support efficient random access and data transmission where potentially a large number of devices would be actively transmitting. The design should leave enough room for the optimizations in both random access and small data transmissions. For massive communication, a promising direction is unsourced multiple access in which the random access and superposition transmission are inherently integrated. This allows uncoordinated parallel transmissions in the same time and frequency resources, so that the latency and device power can be reduced. The number of served users within a given time duration and bandwidth can also be improved.

However, the original unsourced multiple access (at least for the info-theoretical analysis) assumes optimal receiver with joint processing whose computation complexity is too high to be practical. To simplify the problem, the entire process can be broken up into two sub-problems, one for active user detection, channel estimation, and etc. The other for multi-user detection and channel decoding. Under this design principle, a framework, called unsourced sparse multiple access (USMA) can be a candidate solution. The block diagram of its transmitter processing and receiver processing is illustrated in Fig. 1 that consists of two key modules: compressed sensing for preamble generation, and sparse IDMA for main payload transmission.

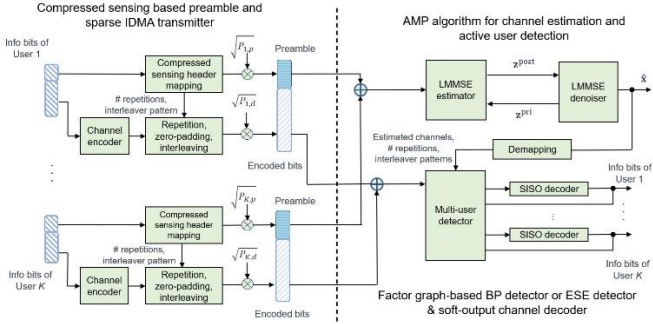

**Figure 1** *Bock diagram of transmitter and receiver of unsourced sparse multiple access (USMA)*

### A. Compressed sensing module

Compressed sensing module serves three main purposes: 1) active user detection, 2) channel estimation, and 3) forming temporal ID. Here the purpose of active user detection is similar to that of preamble detection in traditional random access for eMBB. However, more devices are expected to access the system simultaneously. CS based preamble can efficiently handle such situation than Zadoff-Chu sequence or m-sequence. Note that preambles in traditional random access channels are only for user detections, e.g., demodulation reference signal (DMRS) needs to be additionally allocated for channel estimation, leading to more signaling overhead. In USMA, a temporal ID is constructed by offloading a small number of bits from the data packet to the preamble. These bits can indicate the user-specific patterns of bit interleaver, repetition or zero-padding, to be used for sparse IDMA.

Without losing the generality, let us assume that both base station and device have only one transmit or receive antenna. Also, the channels of all potential users are assumed frequency flat and static within a transmission duration, and can be represented as $\mathbf{x}$ as in Eq. (1). Each channel suffers AWGN, denoted as $n_i$, with zero mean and variance of $\sigma^2$. The basic processing at the transmitter is as follows. First, several information bits are extracted from the packet. These bits form a sequence, e.g., of length $L$, that can be decimally indexed. The index is used to select a column of the compressed sensing matrix $\mathbf{F}_{\text{partial}}$ as in Eq. (1):

$$\mathbf{y} = \mathbf{F}_{\text{partial}}\mathbf{x} + \mathbf{n} = \mathbf{SFx} + \mathbf{n} = \mathbf{Sz} + \mathbf{n} \qquad (1)$$

where $N$ denotes the number of columns of the CS matrix, e.g., $N = 2^L$. The number of rows of the CS matrix, denoted as $M$, corresponds to the number of physical resources, e.g., the preamble length. $\mathbf{y}$ denotes the received signal. Usually, the preamble length is shorter than the number of columns., e.g., $M < N$. The choice of $M$ depends on SNR operating point and some other factors. Ideally, the elements of a CS matrix should be identically independently distributed (i.i.d). Its design needs to balance between the performance and processing complexity. One way to construct $\mathbf{F}_{\text{partial}}$ is to start from a discrete Fourier Transform (DFT) matrix, e.g., $\mathbf{F}$, which is a $N\times N$ matrix. Then, as Eq. (1) shows, a selection matrix $\mathbf{S}$ is used which consists of $M$ randomly selected and reordered rows of an identity matrix. The DFT of the channel vector $\mathbf{x}$ can be represented as $\mathbf{z} = \mathbf{Fx}$ and has efficient implementation such as Fast Fourier Transform (FFT). The number of active devices, denoted as $K_a$, is usually much smaller than $N$, e.g., $K_a \ll N$. Hence, the setting satisfies the notion of compressed sensing, e.g., many elements in $\mathbf{x}$ are zeros. The preamble of $k$-th device is transmitted with the power denoted as $P_{k,\text{p}}$.

At the receiver, approximate message passing (AMP) type of algorithms can be used which generally involve iterative process. In the example of Fig. 1, the AMP algorithm consists of a linear minimum mean squared error (LMMSE) estimator and MMSE denoiser. The essential processing in LMMSE estimator is to take the priori estimate of $\mathbf{z}$, denoted as $\mathbf{z}^{\text{pri}}$, as the input. By considering the priori variance of the estimated $\mathbf{z}$, the posterior estimates of $\mathbf{z}$, denoted as $\mathbf{z}^{\text{post}}$, can be calculated. The processing in MMSE denoiser is simply to calculate the mean and variance of the estimated $\mathbf{z}$. The output of AMP, denoted as $\hat{\mathbf{x}}$, represents the estimated channels of potential users. Certain threshold can be set to decide whether a user is active.

### B. Sparse IDMA module

The design of sparse IDMA supports more superimposed users compared to traditional IDMA. As illustrated in Fig. 1, majority of the payload bits go through a channel encoder, followed by repetitions, zero-padding, bit interleaving, etc. Patterns of repetitions or interleavers are user-specific and can be indicated via the bits carried by the preamble. The encoded bits of $k$-th user are transmitted with the power $P_{k,\text{d}}$.

Its receiver has two key blocks: multi-user detector (MUD) and soft-input-soft-output (SISO) channel decoder. The MUD utilizes the estimated channels $\hat{\mathbf{x}}$,. A parallel MUD is often employed with two typical types. One is the belief propagation (BP) algorithm that operates on factor graphs defined by the repetition, zero-padding or interleaver. While the BP algorithm can exploit the factor graph structure and tends to be more optimal, it is more complex. Alternatively, elementary signal estimator (ESE) can deliver reasonably good performance, yet with significantly less complexity. In ESE, cross-user interference is assumed Gaussian distributed, thus the exact structure of the factor graph can be ignored. Basically, the mean and the variance of cross-user interference can be updated by summing up the estimated means and variances of all other users' signals. For the SISO decoder, maximum a posteriori (MAP) algorithm is often used which can be implemented in different forms, such as BCJR for convolutional codes and BP for low density parity check (LDPC) codes.

### C. Simulation results

Simulations are conducted for Gaussian multiple access channel (GMAC). The packet size is 100 bits and the total number of resources (e.g., modulation symbols) is 30000. The number of active users is up to 300. Here, the size of DFT (e.g., $N$) is set to 8192 to ensure that when $K_a = 300$, per-user collision probability of CS preambles meets the requirement in Table 1, e.g., $1 - \left(1 - \frac{1}{8192}\right)^{300-1} = 3.6\% < 5\%$. While larger values of $N$ can further reduce the collisions, they would cause excessive punctures, thus degrading the performance. The number of payload bits carried by the preamble is $L = \log_2(N) = 13$, about 10% of the entire packet. An intuitive choice of $M$ would be 10% of 30000 which is 3000. However, the optimization also needs to consider the receiver algorithms and the transmit power ratio between the preamble and encoded bits. By extensive trials, the ratio $P_p/P_d$ is set to 10 dB and $M$ is set to 2000, e.g., 6192 rows of the DFT matrix are punctured.

Fig. 2a shows the error probability of CS detector as a function of $E_b/N_0$. It is seen that as $E_b/N_0$ increases to 8 dB, the error rate falls below 1%. The required $E_b/N_0$ slightly decreases as $K_a$ increases. The sparse IDMA module carries $100 - 13 = 87$ information bits. Two coding schemes are simulated: convolutional codes $(133, 171)_8$ and 5G New Radio (NR) LDPC codes, both of code rate = 1/2. The encoded bits are ultimately mapped to 28000 binary phase shift keying (BPSK) symbols. ESE algorithm is employed in MUD. Simulation results of sparse IDMA are shown in Fig. 2b where the performance is measured as the required $E_b/N_0$ for BLER = 5%. In the case of ideal channel estimation, when $K_a$ is below 150, LDPC performs better. However, as $K_a$ exceeds 150, the required $E_b/N_0$ for convolutional codes increases at a slower rate than LDPC codes. Such observation reveals that strong codes for single user case may lose edge over weaker codes when cross-user interference is severe, the block length is short, and iterative receivers are employed. The random coding bound of GMAC of block length = 100 bits at 5% error rate and the performance using convolutional codes with realistic channel estimation are also shown in Fig. 2b. It is seen that the gap to the random coding bound is about 1~1.5 dB. The performance degradation due to the realistic channel estimation is less than 0.1 dB. In addition, the state-of-the-art performance [7] (shown in blue) is compared. Although more advanced MUD, e.g., BP algorithm and stronger codes, e.g., LDPC codes are used, its performance is generally inferior, which may be due to the insufficient overall optimizations of parameters, the combinations of MUD and channel codes, etc.

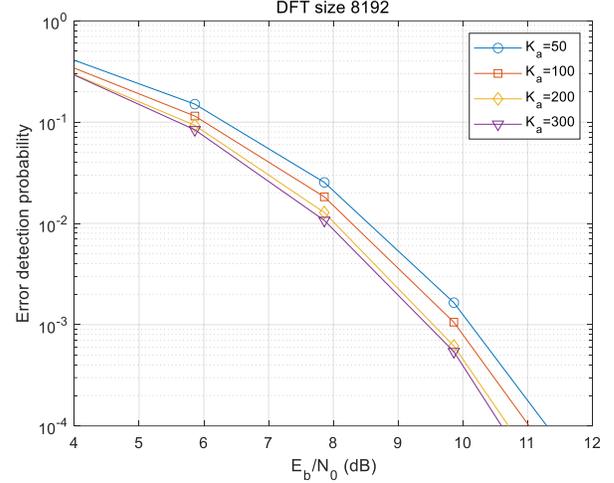

*(a) CS detection performance*

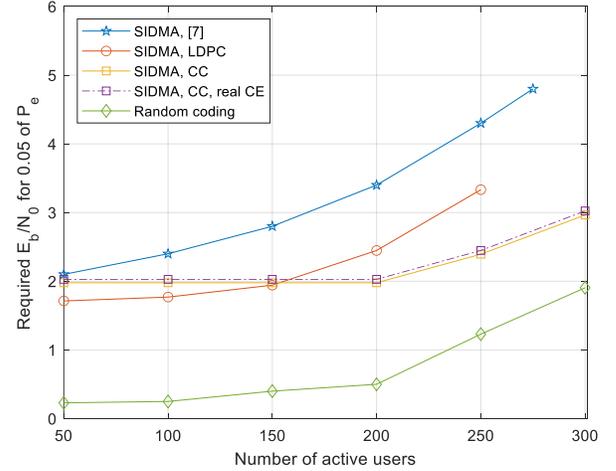

*(b) Sparse IDMA performance*

**Figure 2** *Simulation results of USMA in GMAC channel*

Rayleigh fading with 2 uncorrelated received antennas is also simulated and the BLER results are shown in Fig. 3, assuming ideal channel estimation. It is seen that when $E_b/N_0 = 6$ dB, 1500 active users can be accommodated with BLER < 0.05. Compared to the case of GMAC, a much higher $K_a$ value can be supported. This can be explained by the received power differences between users, which generally benefits MUD. However, the collision probability would be too high, e.g. $1 - \left(1 - \frac{1}{8192}\right)^{1500-1} = 16.7\%$, if $N = 8192$ is still used. This suggests that for fading channels, more resources or design effort should be spent on CS based preambles.

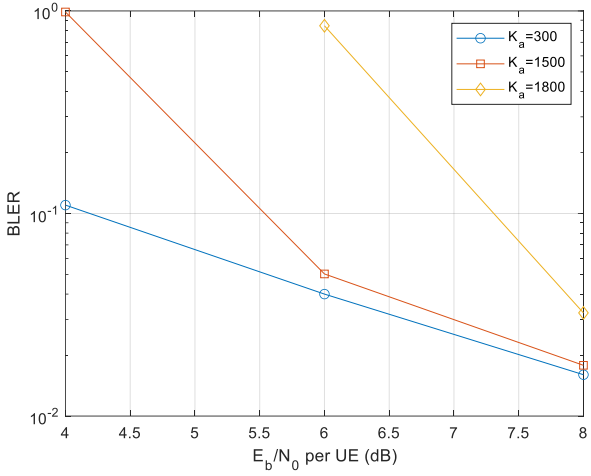

**Figure 3** *Simulation results of sparse IDMA in Rayleigh fading*

## IV. SCALABLE DESIGN FOR AMBIENT IOT

The framework of USMA shown in Fig. 1 is scalable in the sense that: 1) it allows sub-module level optimizations for channel coding, bit interleaver, spreading, compressed sensing matrix, etc.; 2) it can be customized to fit various applications and design parameters of 6G massive communication. In this section, ambient IoT system is elaborated as an example of the scalable design.

### A. Physical layer of ambient IoT

RFID protocol is the traditional air interface of ambient IoT, widely used in warehouse inventory, logistics, etc. It is based on slotted ALOHA mechanism, as shown in the upper part of Fig. 4. In an RFID system, terminal devices and base stations are often called tags and readers, respectively. When operating in 180 kHz bandwidth, the physical layer of RFID is basically a time-division multiple access (TDMA) system. The reader initiates an inventory process by sending the "Query" over the downlink channel. "Query" contains control information for the subsequent uplink transmissions. "Query" is followed by the CW transmission which is also sent by the reader, possibly in the uplink band. After the "Query" is received, each tag would switch to the uplink band and harvest the energy from the CW. With enough number of tags in the system, one or more tags would respond the "Query" within this cycle of inventory by sending a random binary sequence of 16, called RN16. If there is only one tag sending RN16 and the received SNR is adequate, the reader would respond the RN16 by sending acknowledgement (ACK) in the downlink. Once the ACK is received and matches the corresponding RN16, the tag would send the payload which is usually an electronic product code (EPC) of length 96 bits. Simple line code is used to improve the robustness of EPC transmission. If the EPC is successfully received, the reader would send another "Query" for the remaining tags. If multiple tags respond simultaneously, collisions would occur. In this case, mostly likely neither RN16 would be successfully received. Then, a new "Query" would be sent. If there is no response within the inventory cycle, the reader would send another "Query" till it receives a RN16 from a tag. Due to the inability to handle simultaneous transmissions of RN16 or EPC from multiple tags, the resource utilization of RFID is quite low.

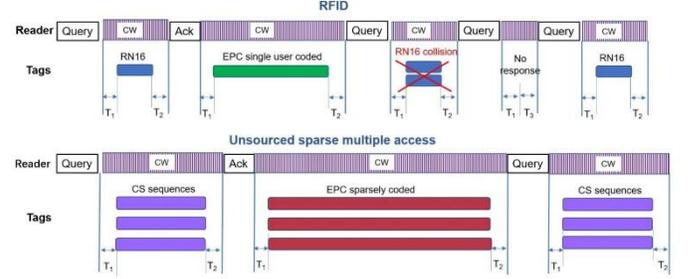

**Figure 4** *Physical layer transmissions of ambient IoT: traditional RFID system vs. USMA*

Spectral efficiency of ambient IoT system can be significantly improved by the customized design of USMA. Its physical layer transmissions are outlined in the lower part of Fig. 4, which can be seen as a variant of the operation in Fig. 1 in the sense that basic mechanism of slotted-ALOHA is kept, e.g., data packets of different users can be sent in different slots, rather than being expanded over a very long frame and transmitted altogether.

With USMA, a reader can detect CS preambles of multiple tags that are trying to access the system simultaneously, as long as their preamble sequences are not the same and the number of active tags in a slot does not exceed certain threshold. After the preamble detection, the reader would decode EPCs of those tags transmitted simultaneously. With these parallel transmissions, the resource can be more efficiently utilized. In addition, the longer transmission time of preamble and EPC in USMA compared to RFID systems can help to extend the coverage.

### B. Specific designs of USMA for ambient IoT

Ambient IoT has very stringent constraints on cost and complexity, especially on the device/tag side. Hence, the following specific optimizations are considered:

#### 1) Compressed sensing matrices

The extremely low cost of tags can only support BPSK or on-and-off keying (OOK) modulation, rendering very limited number of signal levels, e.g., complex-valued signal is not feasible. Hence, instead of using truncated DFT matrix, truncated Hadamard matrix is used for preamble generation.

#### 2) Channel coding schemes

Less complicated coding schemes such as convolutional codes are considered, rather than LDPC codes. At the tag transmitter, convolutional encoding involves only a handful of shift registers' operation. There is no need to store the parity check matrices as for LDPC encoding.

*3) Bit interleavers*

To reduce the memory size, quadratic polynomial permutation (QPP) interleaver is used in which the position of the interleaved bit can be calculated on-the-fly, without the need to look up a stored table. The QPP interleaver is optimized for both CS matrix construction and sparse IDMA to maximize the number of interleaving patterns. The length of QPP interleaver is designed much shorter than the length of CS preamble or EPC sequence, which helps to lower the hardware complexity of Device 1a and Device 1b.

*4) Combined with slotted-ALOHA*

Instead of "expanding" a data packet by 300 times and allowing 300 active users as in Section III-C, smaller scale of "expansion" is considered which can drastically reduce the memory requirements for both CS matrix and sparse IDMA. The system is essentially a *T*-fold slotted-ALOHA where *T* is the maximum number of active users allowed in a slot.

## C. Simulation results of fading MAC

In ambient IoT simulations, each information packet has 96 bits (e.g., EPC length). Considering the typical setting of ambient IoT transmission and the complexity constraint, the maximum value of $K_a$ is set to 15. If the size of Hadamard matrix is set to 256, the collision probability is $1-\left(1-\frac{1}{256}\right)^{15-1} = 5.3\%$ which is close to the target of 5%. Hence, $\log_2(256) = 8$ bits can be used for preamble generation. Considering that $K_a$ can be up to 15, the "expansion" rate is set to about 9 after some trials. That is: 96 - 8 = 88 bits are encoded with convolutional encoder $(133, 171)_8$ of code rate = 1/2, repeated to 376 = 4*(88+6) bits, zero padded, interleaved, then mapped to 800 BPSK symbols. The length of the QPP interleaver is 32 bits. Coded bits are block interleaved.

Regarding the number of rows in the truncated Hadamard matrix (e.g., *M*), as discussed in Section III-C, the compressed sensing module tends to be the bottleneck in fading channels. It means that higher percentage of resources should be allocated for preambles, compared to AWGN channel. In the GMAC simulation in Section III-C, the numbers of resources for preambles and for sparse IDMA are 2000 and 28000, respectively. In ambient IoT simulation, although the percentage of payload bits carried by the preamble is 1/12 of the total (lower than 13/100 in Section III-C), the percentage of resources for preamble is set to 96/896 = 10.7% which is higher than 2000/30000 = 6.7% in Section III-C. The ratio $P_p/P_d$ is set to 10 dB, the same as in Section III-C.

Since the system bandwidth is 180 kHz, the transmission duration of a packet is 896/180k = 5 ms, much shorter than the coherent time of 70 ms when the system operates in 900 MHz and the mobility speed is 3 km/h. Hence, the channel is almost constant during a packet transmission and frequency flat, e.g., single-path Rayleigh fading. The two received antennas are assumed spatially uncorrelated.

Fig. 5 shows the BLER performance of USMA designed for ambient IoT in Rayleigh fading channel of two receive antennas. It is observed that for USMA, as the number of active users is increased from 5 to 15, the required $E_b/N_0$ for BLER = 5% slightly increases, e.g., by less than 0.5 dB. The degradation due to realistic channel estimation is about 1 dB. Single-user case is also simulated where 88 information bits are encoded with the similar R= 1/2 convolutional codes as for USMA, without repetition, zero-padding and interleaving. Its BLER performances of ideal and realistic channel estimation are shown in Fig. 5 which are slightly better than those of USMA of 5 or 15 active users.

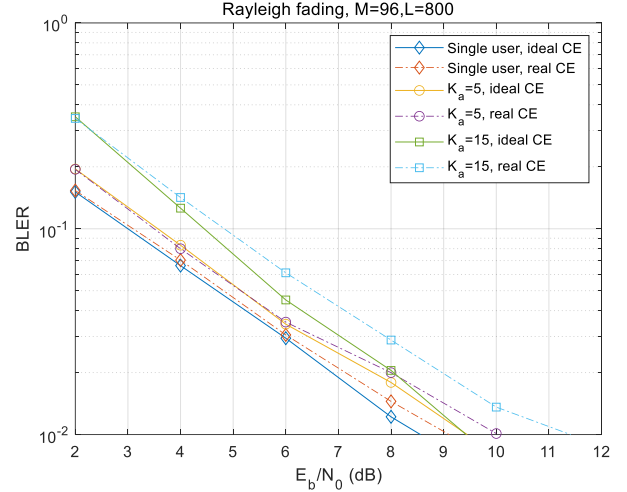

**Figure 5** *Simulation results of USMA customized for ambient IoT under Rayleigh fading channel*

For sparse IDMA, the 88-bit packet spans over 800 BPSK symbols, allowing up to 15 users to be power-domain multiplexed. For traditional RFID system, each user's data (e.g., 96 bits) span over 200 BPSK symbols, which is roughly 1/4 of the resources as of USMA (when the resource overhead of preamble is ignored). However, RFID allows only one user to transmit at a time, meaning that only four users can be time division multiplexed (TDM) within the resources of USMA transmission. Therefore, the capacity or the number of connections in USMA is about 15/4 = 3.75, nearly 4 times of RFID.

Let us also check the preamble part. In traditional RFID system, an RN16 occupies 16 BPSK symbols. In USMA, the CS based preamble spans over 96 BPSK symbols. Assuming 60 RN16 slots are allocated for the slotted-ALOHA access, at most 60*0.367 = 22 tags can be served. With the same number of resources, 60/6 = 10 CS preamble slots can be allocated for USMA. It can be verified that when 90 users randomly pick one of the 10 slots, the probability that more than 15 users simultaneously pick one slot is less than 3%. This means that in terms of resource utilization, the CS based preamble would result in about 90/22 = 4 times efficiency as of RN16.

Another benefit of USMA is that due to the signal "expansion", the average transmit power of either CS based preamble or sparse IDMA based data is much lower than RN16 or EPC packet of RFID, which can help to extend the coverage.

## V. Conclusions

In this paper, the requirements for 6G massive communication and ambient IoT are discussed. A framework of USMA was elaborated that contains compressed sensing based preamble and sparse IDMA. The design optimization includes meticulous choices of parameters, channel coding and receiver algorithms. It was observed that for AWGN channel, in terms of required SNR for certain number of active users, USMA with convolutional codes and a simple multi-user detector can outperform the state-of-the-art scheme of higher complexity. The degradation due to realistic channel estimation is less than 0.1 dB. For fading channel of two received antennas, 5 times number of users can be supported compared to AWGN channel. A customized design for ambient IoT was proposed to further reduce the complexity of USMA. Simulations results showed that compared to traditional RFID system, USMA can provide about 4 times resource efficiency for both random access and data multiplexing.

## Biographies

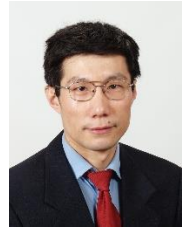

Yifei Yuan was with Alcatel-Lucent and ZTE Corporation He joined China Mobile Research Institute in 2020 as a Chief Expert, responsible for 6G air interface study. He has extensive publications, including 13 books on LTE-Advanced, 5G and 6G. He has over 60 granted US patents

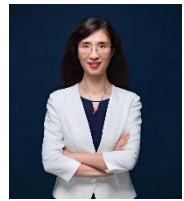

Yuhong Huang is the General Director of China Mobile Research Institute. Over past 30 years, she led many key projects of 2G till 6G.

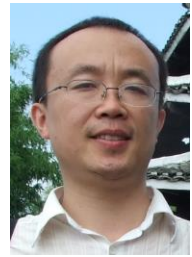

Chunlin Yan has published more than 30 papers on 4G and 5G systems. His current research focuses on MIMO, channel coding, compressed sensing, 6G new multiple access and waveform

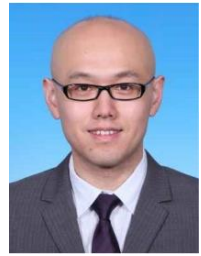

Sen Wang is a chief researcher at China Mobile Research Institute. He holds over 100 patents, and has published over 30 journal and conference papers

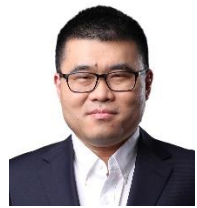

Shuai Ma is an associate director of IoT department of China Mobile Research Institute.

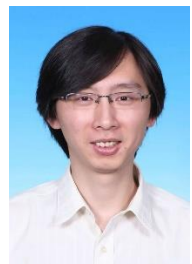

Xiaodong Shen is a chief researcher at China Mobile Research Institute and a delegate of 3GPP.